\documentclass[aip,graphixs]{revtex4-1}

\usepackage{graphicx}
\usepackage{dcolumn}
\usepackage{bm}

\usepackage[utf8]{inputenc}
\usepackage[T1]{fontenc}
\usepackage{mathptmx}

\draft 

\newcommand{\microns}{\, \mathrm{\mu m}}
\newcommand{\nanomet}{\, \mathrm{nm}}

\begin{document}


\title{dark-modes excitation in coupled nano Fabry-Perot structure induced by symmetry breaking} 



\author{Baptiste Fix}
\email[]{baptiste.fix@onera.fr}
\author{Julien Jaeck}
\author{Patrick Bouchon}
\affiliation{DOTA, ONERA, Université Paris Saclay, F-91123 Palaiseau - France}
\author{Nathalie Bardou}
 \affiliation{Center for Nanoscience and Nanotechnology (C2N) - CNRS, Univ. Paris-Sud, Universit\'e Paris-Saclay, 91460 Marcoussis, France}
\author{Benjamin Vest}
 \affiliation{Laboratoire Charles Fabry, 2 avenue Augustin Fresnel, CNRS, Universit\'e Paris-Saclay,  91127 Palaiseau Cedex, France }

\author{Riad Haidar}
\altaffiliation{D\'epartement de Physique, \'Ecole Polytechnique, 91128 Palaiseau, France}
  \affiliation{DOTA, ONERA, Université Paris Saclay, F-91123 Palaiseau - France}


\date{\today}

\begin{abstract}
Absorbing metamaterials have attracted a widespread interest but suffer from the metallic ohmic losses that limits their Q-factor to values around 10. 
In this paper, we report the engineering and experimental characterization of a high Q dark-mode in a metasurface resulting from the coupling of several individual nano-Fabry-Perot structures supporting low-contrast resonances.
The resulting resonance has a Q-factor above 20 and a nearly total absorption. Moreover, we also demonstrate that several dark-mode resonances can be excited simultaneously, by exploiting the coupling of various pairs of nano Fabry-Perot resonators within the same subwavelength period. Finally, the resonance's Q-factor depends monotonously on the thickness of the metasurface and can reach values up to 85.
\end{abstract}


\maketitle 


Absorbing metamaterials have the ability to turn an otherwise reflective surface into a nearly perfect absorber for a given wavelength and polarization. Among those metamaterials, the MIM (metal-insulator-metal) resonator has attracted a widespread interest\cite{cui2014plasmonic} due to its ease of fabrication and the possibility to use large scale fabrication technique such as colloids \cite{moreau2012controlled} or nanoimprint \cite{cattoni2011}. They have been used in various applications, \textit{e.g.,} infrared detection \cite{palaferri2018room}, nonlinear optics, thermal emission \cite{makhsiyan2015shaping}, biosensing \cite{cattoni2011}. 
In most cases, a noble metal is used, which limits the ohmic losses of the structure, and leads to typical Q-factors ( corresponding to the ratio $\frac{\lambda}{\Delta \lambda}$) of 10 in the optical range. For applications taking advantages of light matter interactions or targeting either a narrower spectral band, these losses can be detrimental \cite{khurgin2015deal}.

\begin{figure*}[ht!]
\centering
\includegraphics[width=\linewidth]{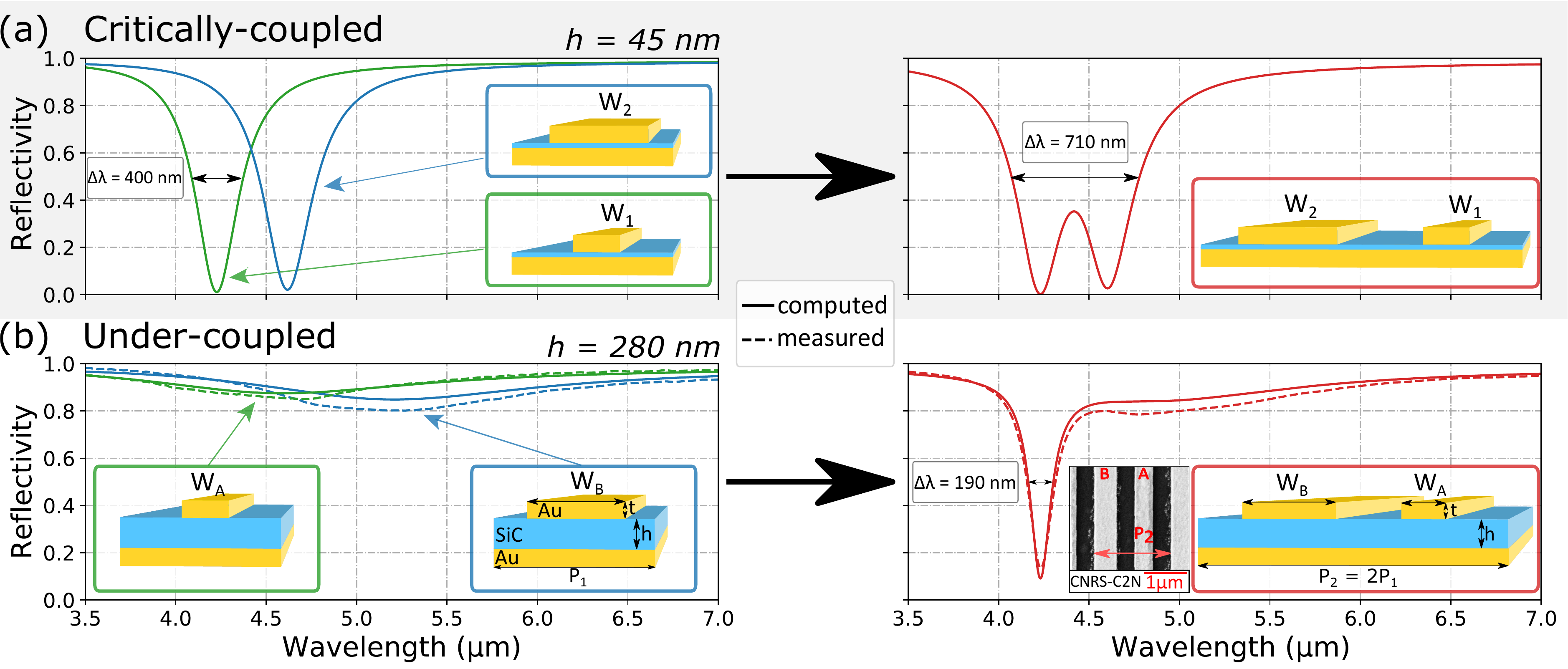}
\caption{(a) Spectral reflectivity under TM-polarised illumination of two standard critically coupled MIM resonators ($h=45 \nanomet$, $W_1 = 475 $ and $W_2 = 525 \nanomet$). Their combination inside a single periodic pattern exhibits two resonances and leads to a widening of the absorbing window. (b) Spectral reflectivity under TM-polarised illumination of two under-coupled Fabry-Perot resonators ($h=280 \nanomet$, $W_A = 400 $ and $W_B = 495 \nanomet$). Their combination inside a single periodic pattern exhibits an unexpected resonance at $\lambda_r = 4.23 \microns$ with a minimum of reflectivity of $14\%$ and a Q-factor of 22.}
\label{fig:spectres}
\end{figure*}

An interesting solution to achieve high Q-factor resonances despite the metallic losses is to engineer the structure in order to enable the excitation of a dark-mode, which, due to its non-radiating nature, has inherently a long lifetime and a high Q-factor. For plasmonic structures, a solution to ensure the existence of a dark-mode is to couple two similar resonators. Thus, the two individual modes hybridize into a symmetric radiant bonding mode and a dark anti-bonding mode\cite{JAIN_2010_dark_coupling}. Such coupled structures have been evidenced for gold nano-particles \cite{Deng_2018_dark_nanoparticle,Sheikholeslami_2010_dark_heterodimer}, nanorods \cite{Gao_18_dark_coupled_dimer,Osberg_2014_dark_antibonding_rod}, grids \cite{Christ_2008_dark_grid,zhou_2011_dark_grid_outofplane} and metamaterials \cite{Omaghali2014_dark_asym_rod}.
Due to their anti-symmetry, dark anti-bonding modes cannot be excited. Yet, it has been demonstrated that creating a grey-mode could enable a free space coupling while maintaining, at least partially, the dark-mode properties. \cite{Omaghali2014_dark_asym_rod,dayal_2016_asym_darkmode,ALNAIB2020_dark_splitring}
Several excitation strategies have been used in the literature to achieve it, such as spatially inhomogeneous \cite{Huang2010_dark_inhomogenous_illumination}, evanescent waves  \cite{Yang_2010_dark_evanescent}, high-angle-of-incidence excitation \cite{zhou_2011_dark_grid_outofplane}. Other strategies have been proposed such as coupling the dark-mode to a bright-mode \cite{Gomez_2013_dark_trimer} or to modify the symmetry of the structure itself \cite{Jeyaram_2013_dark_asym_splitring,Omaghali2014_dark_asym_rod,Cao_12_dark_asym_splitring}.

Therefore, finding a way to use such a dark-mode in a MIM geometry is very appealing to overcome Q-factor limitations experienced by usual Fabry-Perot-like resonances of MIM structures.
However, several studies have been done on symmetry broken MIM metasurfaces, but none have reported the excitation of dark-modes. It is even the other way around, as combination of MIM resonators have been shown to behave nearly independently and used to tailor broadband absorbers \cite{cui2011thin,liu2011taming,Bouchon2012woi, cui2014plasmonic}. 
Here we introduce a general framework for dual MIM to sustain an excitable anti-bonding dark-mode. It is based on the coupling of two undercoupled MIM Fabry-Perot cavities similarly to what was observed with coupled metallic grooves\cite{Fix2017}. This dark-mode is excitable even at normal incidence and we show that the Q-factor value can be increased up to 85 choosing the appropriate configuration. Finally, we demonstrate how several dark-modes can be tailored in the same metasurface.

The MIM resonator geometry, depicted in the insets of Fig~~\ref{fig:spectres}, is made of a stack of a continuous metallic layer, thick enough to be optically opaque, a dielectric layer (thickness $h$) and a patterned metallic layer (thickness $t$). 
Classically, the geometrical parameters are set so that near critical coupling is achieved, \textit{i.e.,} $R\simeq 0$. For instance, Fig~~\ref{fig:spectres}.a shows the computed  spectra of two critically coupled resonators centered at $4.2 \microns$ and $4.6 \microns$ ($W_1=475 \nanomet$, $W_2=525 \nanomet$, $h=45 \nanomet$, $t=50 \nanomet$, period $P_1 = 885 \nanomet$). At resonance, the full width at half maximum (FWHM) of the first resonator is $\Delta \lambda = 400 \nanomet$, which corresponds to a quality factor below 10. 
These two resonators can be combined in the same subwavelength period ($P_2 = 2P_1 = 1770 \nanomet$), leading to two resonances with a broadening of the absorption band ($\Delta \lambda =710 \nanomet$). 
The combined resonators spectrum is well described by the product of the individual resonators spectra
 \cite{koechlin2013analytical}.

In contrast, solid lines on Fig~~\ref{fig:spectres}.b show the numerically simulated spectra of two undercoupled nano Fabry-Perot resonators. 
A resonator is undercoupled when the radiative losses are too high with regards to the internal losses \cite{Joon2011_lightcoupling}.
This is achieved by dramatically increasing the thickness of the SiC layer up to $h=280 \nanomet$, which leads to reflectivity higher than 80\% for both resonator A ($W_A = 400\nanomet$ , $P_1=885 \nanomet$) and resonator B ($W_B = 495 \nanomet$, $P_1=885 \nanomet$).
Surprisingly, when both resonators are combined in the same subwavelength period ($P_2=1770 \nanomet$) to form a new resonator geometry called AB bi-FP, the resulting spectrum is no longer a simple product of the two individual resonators spectra. The reflectivity spectrum shows a blue-shifted resonance near critical coupling ($R=14\%$), with a narrower FWHM ($\Delta \lambda=190 \nanomet$), thus a higher Q-factor of 22.
The optical characterizations of the resonators far-field reflectivity  (see dashed lines on  Fig~~\ref{fig:spectres}.b) confirm the numerical prediction.
Indeed, the agreement between the numerical prediction and the reflectivity spectrum of the AB bi-FP structure for a TM polarised light source with an incidence of $13^\circ$ is very good. The slight mismatch between $4.5 \microns$ to $5.5 \microns$, is attributed to absorption of the SiC layer in this spectral range.

\begin{figure*}[ht!]
    \centering
    \includegraphics[width=\linewidth]{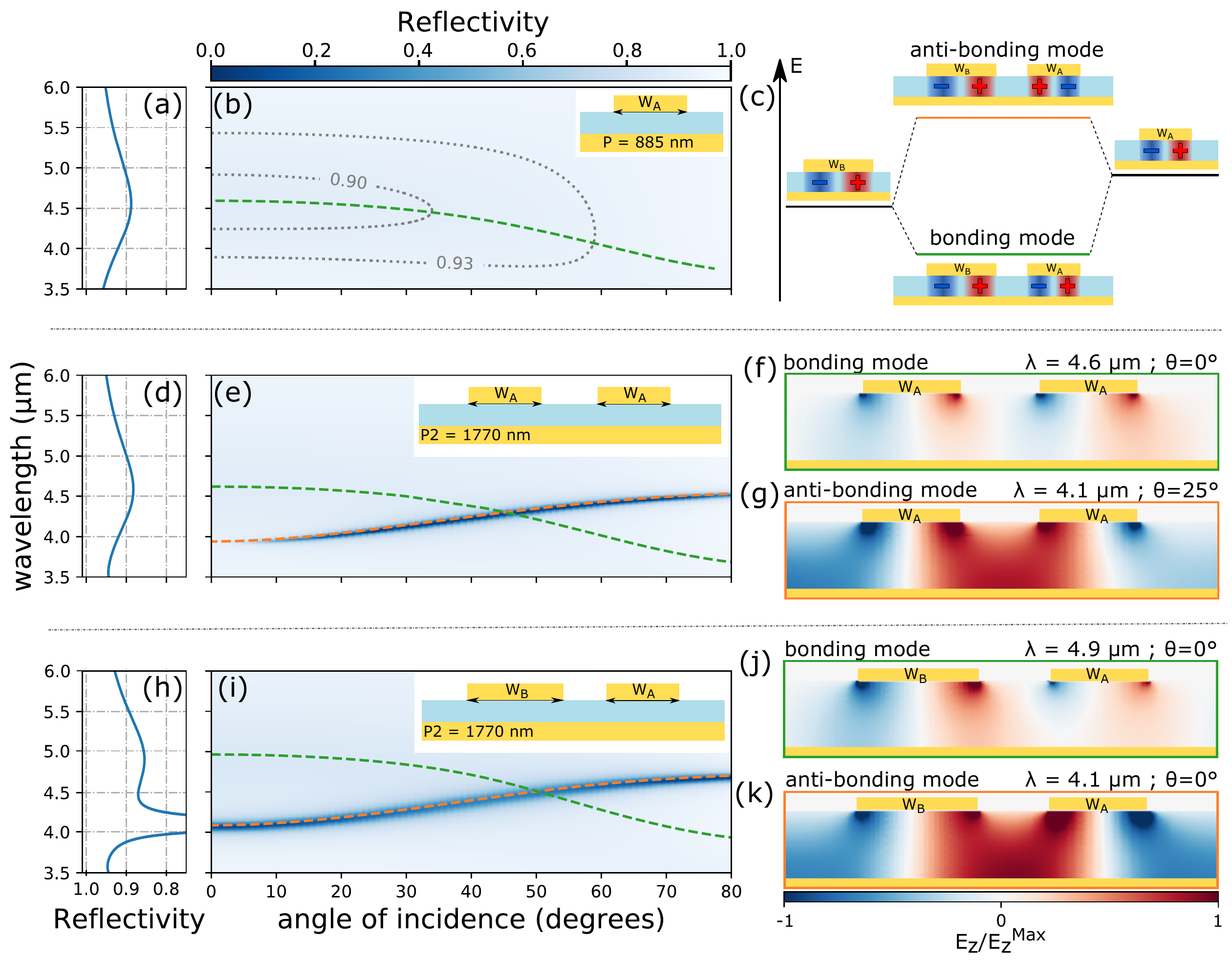}
    \caption{\label{fig:mode_coupling}
        Computed reflectivity spectrum under TM-polarised illumination at normal incidence and with regards to the incidence angle of (a-b) the A single-FP, (d-e) the AA bi-FP and (h-i) the AB bi-FP stuctures. In dashed lines, the computed value of the complex electromagnetic mode (reflectivity poles) of the structures.
        (c) Illustration of the bonding and anti-bonding mode for two coupled nano-FP cavities. 
        (f-g), (j-k) Cartography of the $E_z$ field for the two modes of the bi-FP structures. 
    }
\end{figure*}

In the following of the paper, we discuss the origin, properties and engineering of this narrow resonance in a variety of situations combining individual FP resonators. single-FP, bi-FP, and tri-FP stands for periodic structure with one, two and three ribbons in a given period. The widths of the ribbons are $W_A$, $W_B$ and $W_C$ and for the sake of clarity, the single-FP are named A, B and C, the bi-FP are named AB and BC and the tri-FP ABC. 
In the specific case where two ribbons are identical, but not equally spaced, the structure is named AA bi-FP structure.
The single-FP structures have a period $P_1 = 885\nanomet$ while the period of the bi-FP structures is $P_2 = 2P_1$ with the ribbons evenly distributed in the period.

We now discuss the origin of the narrow resonance of the AB structure on Fig~~\ref{fig:spectres}.b.
The complex electromagnetic modes of the A single-FP structure, and the AA and AB bi-FP structures have been calculated as a function of the illumination angles (see dashed lines in Fig~\ref{fig:mode_coupling}.b,e,i).




These modes have been super-imposed with the structures reflectivity spectra at these illumination angles. In the case of the single-FP structure, only the usual bright gap-plasmon (green dashed line) mode exist and correspond to a highly under-coupled resonance ($R_{min}\simeq $ 90\%), as shown in Fig~\ref{fig:mode_coupling}a).

For the bi-FP structures, two modes exist and correspond respectively to the bonding and anti-bonding mode. They result from the hybridization of the resonances of individual FP resonators combined in a given period (see Fig~\ref{fig:mode_coupling}.c). 
A numerical search of the reflectivity poles of the metasurface allows to identify these two modes. In the case of AA bi-FP structure, the bonding, lower-energy mode (green dashed line on Fig~\ref{fig:mode_coupling}.e), provides a low-Q resonance centered around 4.5 um at normal incidence (see Fig~\ref{fig:mode_coupling}.d), corresponding to the single-FP resonance, blueshifting at higher incidences. The mode exhibits a symmetric field with respect to the middle plane of a period (see Fig~\ref{fig:mode_coupling}.f).

In contrast, the anti-bonding, higher-energy only exhibits a narrow reflectivity drop at non-zero incidence (orange dashed line on Fig~\ref{fig:mode_coupling}.d). This is a direct consequence of the antisymmetric field distribution of this mode, which make it dark under a symmetric, normal incidence excitation (see Fig~\ref{fig:mode_coupling}.g).


Interestingly, symmetry breaking induced by oblique excitation allows to couple far field radiation into the mode volume.
The resulting resonance exhibits a high Q-factor of about 50 to be compared with the bonding mode resonance with a Q-factor around 10.

In the case of the AB bi-FP structure experimentally studied in the Fig~\ref{fig:spectres}.b, the modification of one ribbon width break the middle plane of symmetry (Reflectivity spectra, map and mode structures are depicted on Fig~\ref{fig:mode_coupling}.h-k). This enables the excitation of the "dark-mode" (now strictly a grey-mode) at any incidence angle. The increased coupling between the anti-bonding mode and the free space decreased the resonance Q-factor compared to the case of the "purely dark-mode" of the AA bi-structure. Still, the Q-factor of the resonance is about 25 which is still 2 to 3 times higher than the gap-plasmon resonances.

\begin{figure}[ht!]
    \centering
    \includegraphics[width=\linewidth]{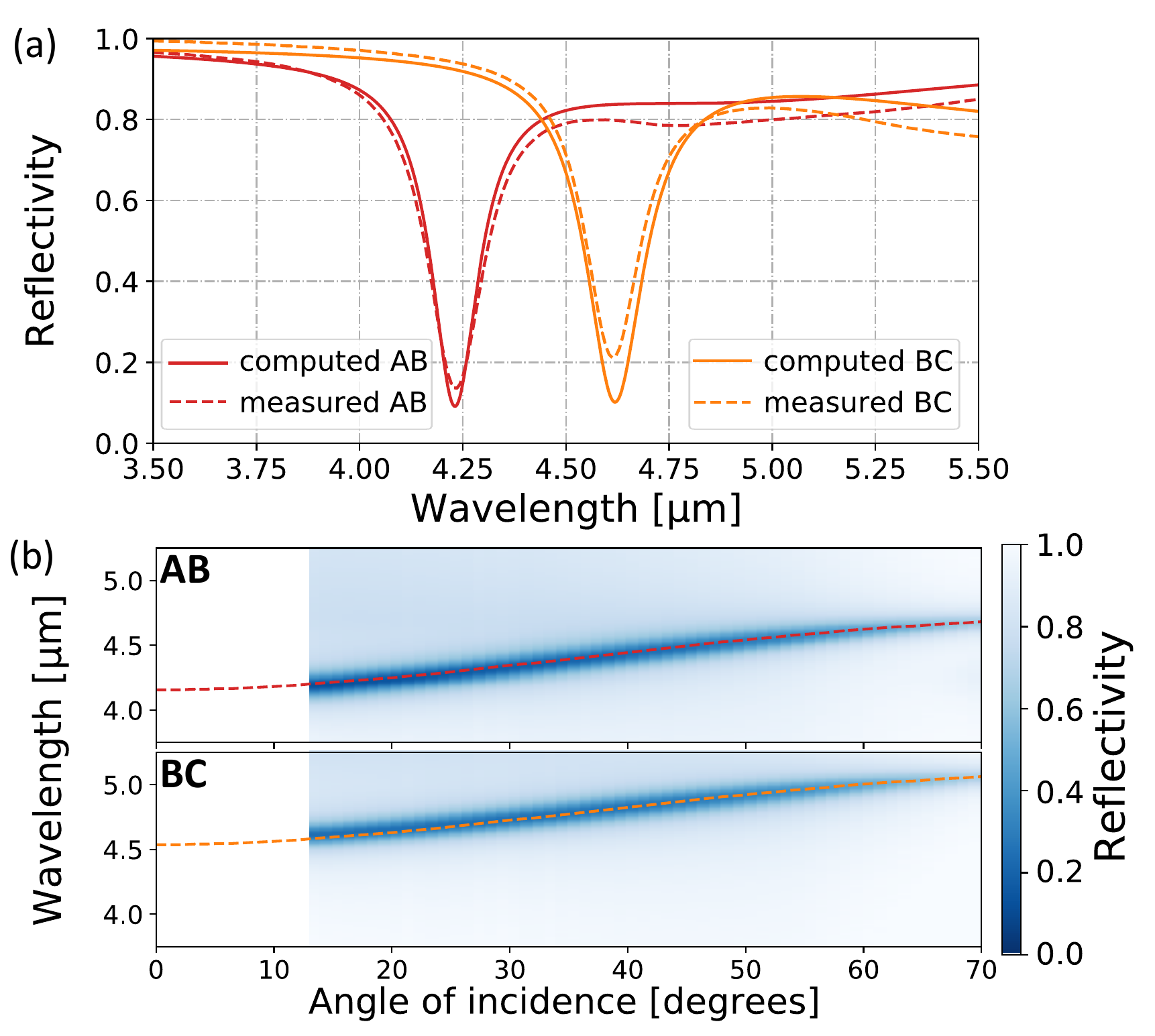}
    \caption{\label{fig:AB-BC}(a) Measured (dashed) and computed (solid) reflection spectra of the AB (red) and BC (orange) bi-FP structures at $13^\circ$ of incidence. The resonance wavelength of the bi-FP structure is tunable by the modification of one or two ribbon widths. ($\mathrm{W_A}$=400$\nanomet$,$\mathrm{W_B}$=495$\nanomet$,$\mathrm{W_C}$=600$\nanomet$) (b) Measured reflectivity map of the AB and BC bi-FP structure for a TM polarised illumination between $13^{\circ}$ and $70^{\circ}$ of incidence. The dotted lines represent the evolution of the resonance wavelength obtained by the numerical simulation.}
\end{figure}

The anti-bonding resonance has three appealing properties. First, the resonance wavelength can be tuned by adjusting the dimensions of the pattern while maintaining the critical coupling. Secondly, it is possible to introduce more anti-bonding resonances by adding under-coupled FP cavities inside the periodic pattern. Thirdly, the quality factor can be tuned independently from the resonance wavelength.

The Fig~\ref{fig:AB-BC}.a demonstrates the tunability of the resonance: it compares the numerically simulated and experimentally measured reflectivity spectrum at $13^\circ$ of incidence of the AB and BC bi-FP structures, the two structures only differing by the width of one of their two ribbons. Both spectra have a similar shape, with a critically coupled resonance with $14\%$ (resp $21\%$) reflectivity and a Q-factor of 22 (resp 25) but at two different resonant wavelengths $\lambda_r = 4.23 \microns$ and $\lambda_r = 4.61 \microns$.
The measured reflectivity diagram as a function of the angle of incidence is depicted in Fig~\ref{fig:AB-BC}.b for TM-polarised illuminations for the AB and BC bi-FP resonators. The dotted lines represent the computed resonant wavelengths. Both resonators have a similar angular behavior, mostly insensitive to the angle of incidence below 20 degrees, and slightly red-shifting above. 

\begin{figure}[ht!]
    \centering
    \includegraphics[width=\linewidth]{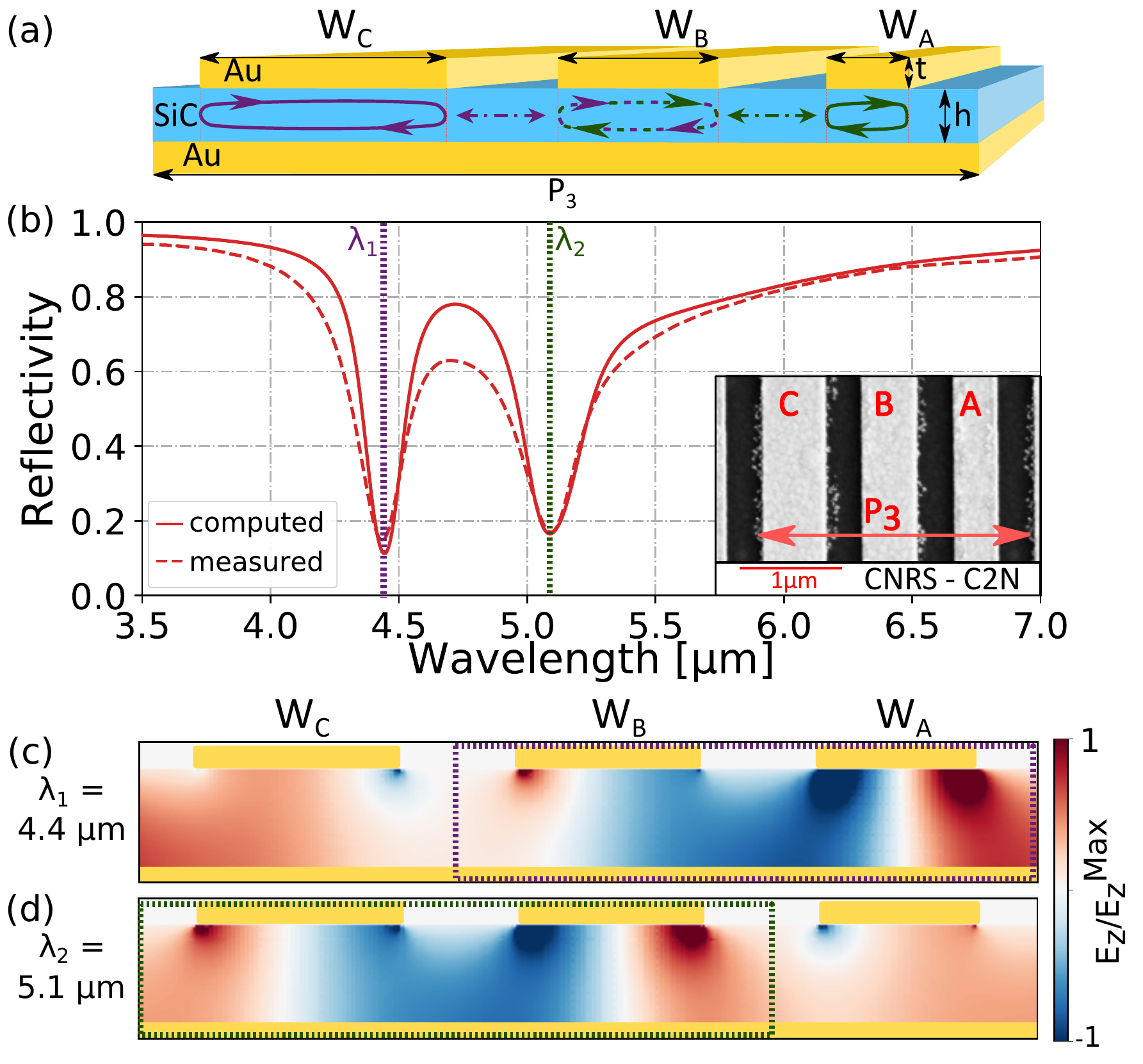}
        \caption{\label{fig:ABC}(a) Scheme of the ABC coupled-FP structure with period $P_3 = 3P_1$, $\mathrm{W_A}=405 \nanomet$, $\mathrm{W_B}=500 \nanomet$, $\mathrm{W_C} =600 \nanomet$. (b) Measured and computed reflection spectra of the ABC coupled-FP structure at $13^{\circ}$ incidence. 
        Two resonances are located at $\lambda_{1} = 4.43 \microns$ and $\lambda_{2} = 5.12 \microns$ with minima of reflectivity of $15\%$ and Q-factors of 15 and 11 respectively. The inset is an SEM image of the fabricated structure.
        Computed distribution of the $E_{z}$ field at (c) $\lambda_{1} = 4.43 \microns$ and (d) $\lambda_{2} = 5.12 \microns$.
        }
\end{figure}

It is possible to combine even more under-coupled nano Fabry-Perot inside the same sub-wavelength period. As just seen, in Fig~\ref{fig:AB-BC}.a, the B single FP can be coupled to two different FP (namely A and C), leading to two bi-FP resonators with different resonance wavelengths.
 Therefore, three different ribbons can be placed in the same subwavelength period giving birth to a tri-FP structure, as illustrated in Fig~\ref{fig:ABC}a.
This structure has been fabricated on the same sample than the previously studied resonators, so the individual FP cavities would still be under-coupled. A top view SEM image of the tri-FP is shown in the inset of Fig~\ref{fig:ABC}b.
The computed and measured reflectivity spectra of the ABC tri-FP structure are plotted in Fig~\ref{fig:ABC}b. The experimental characterization of the resonator reflectivity shows two resonances at $\lambda_{1} = 4.43 \microns$ and $\lambda_{2} = 5.12 \microns$ with a lower-than-$20\%$ reflectivity in both cases and quality factor $Q_1=15$ and $Q_2=10$. These quality factors are lower than expected theoretically for a plane wave excitation ($Q_1=28$ and $Q_2=22$), which is due to the focusing of the illumination.
In order to reach a good agreement between the experiment and the computation, a $\pm 6^{\circ}$ aperture of the focused beam inside the FTIR has to be taken into account. The remaining 20\% reflectivity difference between 4.5 $\microns$ and 5 $\microns$ can be attributed to the SiC absorption. 
The ABC tri-FP spectrum is not the mere combination of the two bi-FP (AB and BC) responses of Fig~\ref{fig:AB-BC}. 
Indeed, the two resonances are red-shifted with comparison to their bi-FP counterparts.
It can be explained by looking at the computed cartography of the field at the two resonances (Fig~\ref{fig:ABC}.c,d). We can locate on the two maps the field distribution of the anti-bonding modes resonances of AB or BC. They display a stronger field amplitude unbalance than the corresponding resonances of Fig~\ref{fig:mode_coupling}.k. It can be inferred from this comparison that the presence of the third resonator affects the anti-symmetry of the field distribution at resonance, thus further "lightening" the initially dark mode.

\begin{figure}[ht!]
\centering
\includegraphics[width=\linewidth]{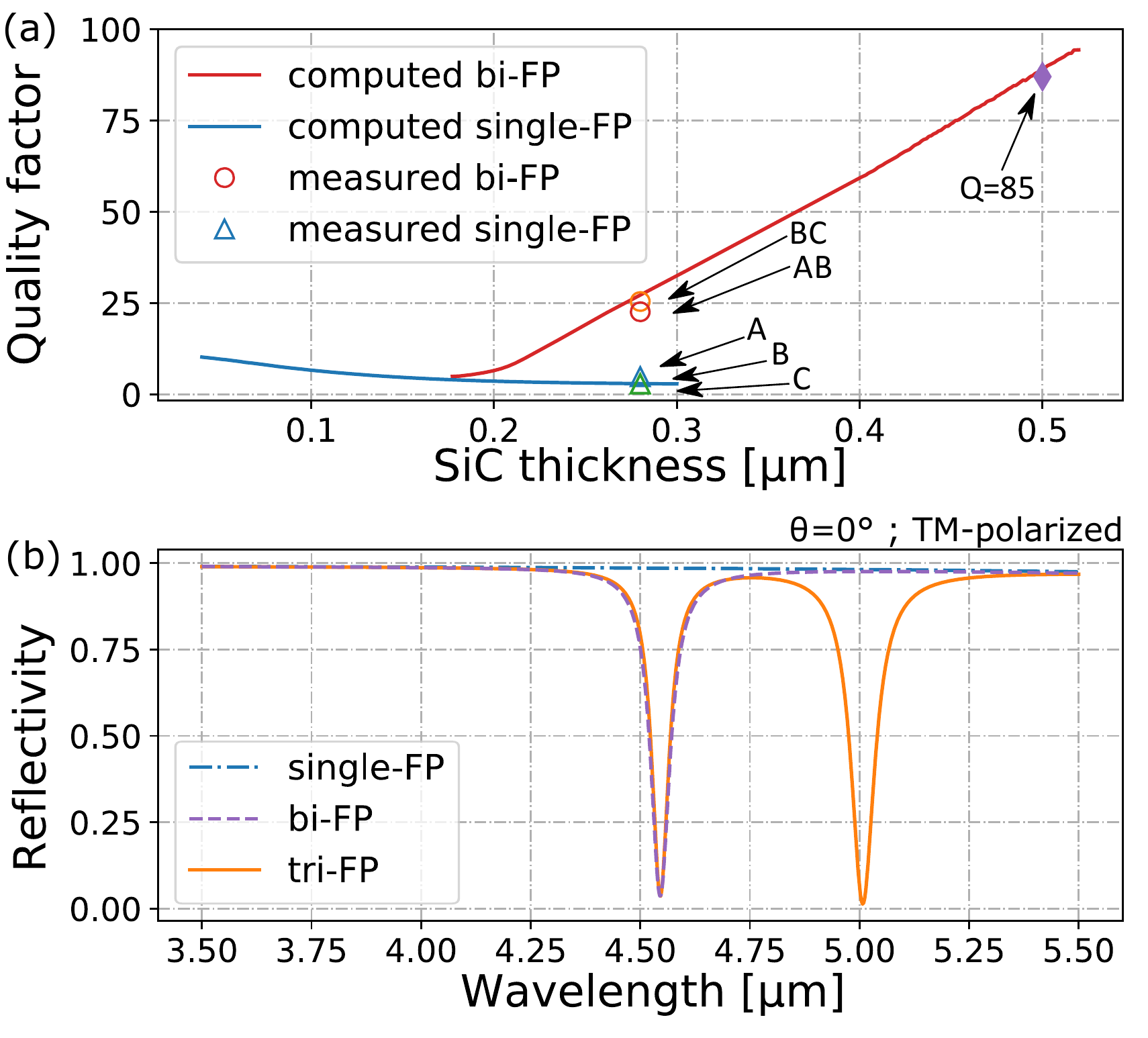}
\caption{\label{fig:interet} (a) Computed evolution of the Q-factor with regards to the SiC layer thickness for the AB bi-FP (red) and B single-FP (blue) structures. The measured Q-factors of the AB, BC bi-FP structures and the A,B and C single FP-structures,reported on the graphic, are in good agreement to the simulation and prove the high Q-factor accessible for the bi-FP structure. (b) Reflectivity spectrum of a bi-FP ($W_A= 450 \nanomet$ and $W_B=610 \nanomet$) and a tri-FP ($W_A= 260 \nanomet$, $W_B = 470 \nanomet$ and $W_C=630 \nanomet$) with a higher SiC thickness ($h=500 \nanomet$). As expected, the quality factor of the resonance is about 85 for all the resonances.}
\end{figure}

In the following, we explore the possibility of adjusting the Q-factor to achieve higher values.
Fig~\ref{fig:interet} shows the calculated evolution of the Q-factor of single-FP and bi-FP resonances as a function of the dielectric layer thickness. The bi-FP resonance exists for a thickness above 180$\nanomet$, at which point the individual FP are undercoupled.From this minimum thickness, the Q-factor of the bi-FP resonance increases linearly up to values as high as 85 for a $500\nanomet$-thick SiC layer.
The experimental Q-factors measured on the different nanostructures are reported on the Fig~\ref{fig:interet}a and are in good agreement with the numerical simulation, the small difference being here again due to the aperture of the focused incident beam inside the FTIR. Fig~\ref{fig:interet}b shows the example of one bi-FP and one tri-FP structure for a $h=500 \nanomet$ thick SiC layer. Compared to the previous structures, the ribbons widths had to be re-evaluated to 450 nm and 610 nm for the bi-FP and 260 nm, 470 nm and 630 nm for the tri-FP in order to maintain the resonances around $4.5\microns$ and $5 \microns$.
Moreover, since this property rely only on the insulator thickness while the resonance wavelength is mostly driven by the ribbons width, a resonator is easily design to exhibit a resonance at the desired wavelength and Q-factor.

To conclude, we have demonstrated that the engineering of the coupling between two nano FP-like nanotructures when using symmetry breaking enables the excitation of a high Q-factor anti-bonding "dark-mode". 
We have confirmed those predictions experimentally and observed anti-bonding resonances with Q-factors as high as 20 for a set of resonators designed to provide various resonant wavelengths near 4.5 microns.
Moreover, we have demonstrated that 
nano-FP like structures can be combined within a same subwavelength period to provide several anti-bonding resonances.
Finally, additional computation show that the structure Q-factor is tunable through the thickness of the insulating layer and can reach values up to 85 for both bi-FP and tri-FP structures.
Thanks to their simple fabrication process and their high Q-factor compared to other MIM structures, we expect these nano-resonators to be useful for thermal emission, chemical sensing or low dimension spectral filter.

\textbf{Method}
The numerical simulation of these nanostructures are performed with the B-spline modal method which allows fast and exact electromagnetic analysis on a non-uniform mesh.\cite{bouchon2010bmm}. 
The gold is modeled using a Drude formula\cite{palik1985handbook} 
: $\epsilon\left(\lambda\right) =  1 - \left[ (\lambda_p/\lambda + i\gamma ) ~~ \lambda_p/\lambda \right]^{-1}$ with $\lambda_p =  158.9 \nanomet$ and $\gamma = 0.0077 $  while the SiC is modeled with a real permitivity $\epsilon = 7.29$.
The complex electromagnetic modes of the structures have been calculated using the RCWA method \cite{Hugonin2005ret}.
The nanostructures have been fabricated on a 2mm x 2 mm footprint area on the same sample. 
The bottom continuous gold layer (200 nm thick) was deposited by electron-beam evaporation on a silicon wafer. 
Then, a 280 nm thick layer of SiC was deposited by evaporation. 
The patterns were realized by electron-beam lithography on a Ma-N 2403 resist follow by the deposition of 2 nm of titanium and 50 nm of gold.
Finally a liftoff was performed using acetone. 
Using the Scanning Electron Microscope (SEM), we measured the dimensions : $P_2=1770\nanomet$ and $\mathrm{W_A}$=400$\nanomet$,$\mathrm{W_B}$=495$\nanomet$, $\mathrm{W_C}$=600$\nanomet$.
The ABC tri-FP structure geometrical parameters have been measured using the SEM as : $\mathrm{W_A}=405 \nanomet$, $\mathrm{W_B}=500 \nanomet$, $\mathrm{W_C} =600 \nanomet$. 
The optical characterizations were achieved with a Bruker Vertex 70V Fourier transform infrared (FTIR) spectrometer.
This research was supported by a DGA-MRIS scholarship.


\bibliography{biblio.bib}

\end{document}